\documentclass[proof]{pasj00}

\SetRunningHead{N.Kawano, A.Ohto, and Y.Fukazawa}
{Chandra X-Ray Spectral Analysis of 2A 0335+096 and Abell 2199}
\Received{2002/08/27}
\Accepted{2003/03/24}
\Published{2003/06/25} 

\begin{document}
\title{Chandra X-Ray Spectral Analysis of Cooling Flow Clusters,\\ 
2A 0335+096 and Abell 2199}

\author{Naomi \textsc{Kawano}, Akimitsu \textsc{Ohto}, 
and Yasushi \textsc{Fukazawa}}

\affil{Department of Physical Sciences, School of Science,
Hiroshima University, }
\affil{1-3-1 Kagamiyama, Higashi-Hiroshima, Hiroshima 739-8526}
\email{kawano@hirax7.hepl.hiroshima-u.ac.jp}


\maketitle

\begin{abstract}
We report on a spatially resolved analysis of Chandra X-ray data 
on a nearby typical cooling flow cluster of galaxies 
2A 0335+096, together with A 2199 for a comparison.
As recently found in the cores of other clusters, 
the temperature around the central part of 2A 0335+096 
is 1.3--1.5 keV, which is higher than that inferred 
from the cooling flow picture.
Furthermore, the absorption column density is almost constant 
against the radius in 2A 0335+096; there is no
evidence of excess absorption up to 200--250 kpc.
This indicates that no significant amount of cold material, 
which has cooled down, is present.
These properties are similar to those of A 2199.
Since the cooling time in the central part is much shorter 
than the age of the clusters, a heating mechanism, 
which weakens the effect of radiative cooling, is expected 
to be present in the central part of both clusters of galaxies.
Both 2A 0335+096 and A 2199 have radio jets associated 
with their cD galaxy.
We discuss the possibility of heating processes caused by 
these radio jets by considering the thermal conduction 
and the sound velocity together with the observed disturbance 
of the ICM temperature and density. 
We conclude that the observed radio jets can produce 
local heating and/or cooling,
but do not sufficiently reduce the overall radiative cooling.
This implies that much more violent jets, whose emission has 
now decayed, heated up the cooling gas $>10^9$ years ago.
\end{abstract}

\section{Introduction}
Many clusters of galaxies exhibit a sharp profile around the
X-ray emission peak at the center, and
their intracluster medium (ICM) becomes cooler toward their center.
This is apparently consistent with 
the idea of a cooling flow model (review; \cite{fabian94}).
However, other observational phenomena have also been found 
and cannot agree with the simple
cooling flow picture (\cite{makishima01}; \cite{tamura01a}).
In order to obtain more information about the physical condition of 
the ICM in the central part of clusters, 
spatially resolved measurements of the temperature, 
absorption column density, and metal abundance are needed.
The spatial resolution and effective area of previous X-ray
satellites did not allow us to resolve the innermost cluster 
regions in detail.

Recent observations with Chandra and XMM-Newton revealed
complex physical properties of the ICM in the central region 
of the Perseus cluster (\cite{fabian00}; \cite{schmidt02}), 
Hydra-A cluster (\cite{mcnamara00}; \cite{david01}), A 1795 
(\cite{tamura01a}; \cite{fabian01}; \cite{ettori02}), 
A 1835 (\cite{schmidt01}; \cite{peterson01}), 
and A 496 (\cite{tamura01b}) clusters,
which are typical cooling flow clusters.
One of the important results is that the ICM temperature 
at the central region is not as low as $\sim$1 keV, 
which is expected from the simple cooling flow model.
X-ray emission of the Perseus and Hydra-A clusters shows 
holes with X-ray bright rims, which are caused by an ICM 
displacement by the radio jet. 
These results indicate that the radio jet is violently interacting 
with the ICM, and is thus a possible candidate of heating sources 
which would reduce the cooling flow.
However, the ICM temperature of bright rims surrounding 
the radio jet in the Perseus cluster is
$\sim$ 2.7 keV, which is cooler than the ambient ICM.
In the case of the Hydra-A cluster,
there is no correlation between the radio jets 
and spectral hardening due to shock of radio jet. 
On the other hand, evidence of jet heating is reported
for MKW3s (\cite{mazzotta02}). 
The region of depression in the X-ray surface brightness, 
which lies south of the X-ray peak, 
exhibits a higher temperature than the surrounding one, 
indicating energy injection from the central radio galaxy.

In this way, two contrasting results coexist concerning 
whether the heating mechanism associated with radio jets 
exists or not.
Therefore, it is important to obtain much more information 
on the X-ray properties in the central cluster region.
2A 0335+096 ($z=0.035$) and A 2199 ($z=0.0305$) are 
nearby X-ray bright cooling flow clusters of galaxies.
The cooling flow rates are relatively large, 
181 M$_{\odot}$ yr$^{-1}$ (\cite{edge92}) and
90 M$_{\odot}$ yr$^{-1}$ (\cite{owen98}), 
for 2A 0335+096 and A 2199, respectively.
They are accompanied by radio jets at their center, 
especially for A 2199, whose radio jets from the cD galaxy 
NGC 6166 seem to interact with the ICM (\cite{owen98}).
It is thus possible that heating or cooling by the radio jets
takes place in the central region of both clusters.
According to these properties, they are good targets to approach 
the problems on the cooling flow, in terms of the temperature 
structure and excess absorption. 
They are bright enough to be resolved finely 
by the Chandra observatory with good photon statistics. 
The spectral coverage provided by Chandra allows
an accurate estimate of their ICM temperature, 
which is expected to be around 4 keV. 
Significant excess absorption was reported for 2A 0335+096 
by \citet{white91} based on the Einstein SSS observation; 
they claimed that it is attributed to cooling flow.
The low galactic absorption toward A 2199 is also advantageous
to constrain excess absorption. 
The high spatial resolution provided by Chandra
allows us to spatially resolve the X-ray emission of 
these clusters on the arcseconds scale. 

We report in this paper on spatial and spectral
analyses of 2A 0335+096 and A 2199, both observed 
with ACIS-S on board Chandra.
The results of an analysis of the Chandra observation of A 2199 
was presented by \citet{johnstone02}, who derived several 
X-ray properties of the central cooling region.
Since these two clusters are a good pair for a comparison 
to understand the central cooling region of clusters of galaxies, 
as described above, we also reanalyzed the Chandra data on A 2199 
in the same way as for 2A 0335+096.
Throughout this paper, we assume $H_0 = 50$ km s$^{-1}$ Mpc$^{-1}$;
$1'' = 1.02$ kpc for 2A 0335+096
and $1''= 0.89$ kpc for A 2199, respectively.

\section{Observation}
We observed 2A 0335+096 on 2000 September 16, 
with a back-illuminated CCD chip, ACIS-S3, loaded onto Chandra. 
The total exposure time was 20 ks (obsid 919).
A 2199 was also observed with the Chandra ACIS-S3 
on 1999 December 11.
For A 2199, we used archival data (obsid 498), 
and the total exposure time was 19 ks. 
When we started the analysis, no blank sky data were available, 
and thus we employed data of the galaxy NGC 3184 as background, 
since it was observed in the same ACIS set up as both clusters, 
and does not contain any bright X-ray sources.
We confirmed that the background level of the NGC 3184 data is
consistent with that of on-source data. 
The background was estimated in the same detector region 
as on-source data by using the software $acisbg$ 
which is provided by Maxim Markevitch
\footnote{$http://hea-www.harvard.edu/~maxim/axaf/acisbg/$}. 
All of the data retrieved from the Chandra Data Center were 
at level 1--2.
We used the CIAO 2.0 software package provided by CXC 
for the data analysis, and all spectra were corrected 
with the calibration product CALDB 2.9. 

\section{Analysis and Results}

\subsection{X-Ray Imaging Analysis of 2A 0335+096}
The X-ray image of 2A 0335+096 in the 0.4--9.0 keV band 
is shown in figure 1a. 
The X-ray emission of 2A 0335+096 does not seem to be 
circularly symmetric; the south region of the center is brighter 
and the NE region is relatively fainter. 
A hole is seen in the direction of NW at $40''$ 
of the cluster center. 
This hole is not associated with any radio sources, 
and is thus thought to be the same feature as `ghost' cavities
reported for the Perseus cluster and Abell 2597
(\cite{churazov00}; \cite{mcnamara01}), 
which are considered to be fossils of radio-bright cavities 
after electron cooling. 
The VLA radio observation of 2A 0335+096, performed 
in 1987 and 1992, revealed that it has twin jets
which emanate from the cD galaxy, and are extended up to $\sim10''$
toward the NE and SW  (\cite{sarazin95}). 
The radio image in the central region of 2A 0335+096 
is shown in figure 1b. 
A spatial resolution of 16$''$ did not allow us to resolve 
finer structures comparable to the ones observed in Chandra. 
For the first time, however, we noticed a weak correlation 
between the X-ray emission and the radio jet in the direction 
of NE from the center, 
thanks to the high spatial resolution of Chandra.
On the other hand, \citet{johnstone02} 
found that the X-ray emission of A 2199 is displaced 
by the radio jet at the center more clearly than 2A 0335+096.
A short radio jet extending up to $\sim3$ kpc is visible
on the E and W sides of the cD galaxy.

In order to derive physical quantities of the ICM, such as 
the gas density, the X-ray surface brightness profile 
was fitted by the canonical $\beta$-model (\cite{jones84}), 
and is shown in figure 2.
Here, the cluster center was determined to be the position of 
the cD galaxy, which almost coincides with the root of radio jets. 
The best-fit parameters and reduced chi-square are listed 
in table 1.
The profile is rather well-fitted by a single $\beta$ model;
there is no excess in the X-ray surface brightness at the center.
The $\beta$ and core radius, $r_{\rm c}$, are consistent 
with $\beta=0.55$ of the ROSAT HRI observation (\cite{sarazin92}) 
and $\beta=0.60\pm0.05$ and $45''\pm15''$ of the ASCA GIS 
(\cite{fukazawa97}).
Therefore, the X-ray surface brightness of 2A 0335+096 can be 
quite well represented by a single $\beta$-model 
from the most inner to outer region.

\subsection{Radial Profiles of the Spectral Properties for 2A 0335+096}
Although the ICM distribution of 2A 0335+096 is not 
circularly symmetric, as described in subsection 3.1,
we first investigated the spectral variation toward 
the radial direction in order to understand the overall 
features of the ICM. 
To obtain radial profiles of the temperature, 
column density and metal abundance, 
spectral fittings were performed for several annuli centered 
on cD galaxies in the 0.5$-$10 keV region. 
The background spectra and response matrices were extracted 
from the corresponding region. 
The background-subtracted spectra are shown in figure 3a.
Here, we pick up four characteristic spectra at
$0''-10''$, $30''-40''$, $60''-90''$, and $150''-180''$ annuli. 
Fe-L lines become stronger 
and shift to the lower energy side as the radius becomes small, 
indicating that the temperature decreases
toward the cluster center. 
We also show the spectra of A 2199 for comparison in figure 3b.
It can be clearly seen that the absorption 
in the lower energy band is quite different 
between two clusters.
This is mainly due to a large difference in the galactic column
density: 1.7$\times10^{21}$ cm$^2$ and 8.7$\times10^{19}$ cm$^2$ 
for 2A 0335+096 and A 2199, respectively. 
This well demonstrates that the ACIS-S spectrum is
very sensitive to the photoelectric absorption.

These spectra of 2A 0335+096 were fitted with a one-temperature 
MEKAL thermal-plasma model (\cite{liedahl95}) 
with photoelectric absorption.
The free parameters are the temperature ($kT$), 
column density ($N_{\rm H}$), metal abundance ($A_{\rm Fe}$) 
and normalization. As a result, 
the reduced chi-squares, $\chi^2$/dof, are 1.1--1.6 for radii 
beyond 40$''$, and 1.6--2.7 within 40$''$.
We then tried to fit the spectra with the two-temperature
MEKAL model. 
The abundance is tied to be the same among two components, 
and the temperature of the hot component is fixed at 3.3 keV, 
which is that of the outermost region.
The fittings were improved for spectra within 40$''$ from the
center, and the reduced chi-squares $\chi^2$/dof became smaller 
by 0.3--0.5. The $\chi^2$/dof of one and two-temperature fittings 
in the center of 2A 0335+096 are shown in table 2. 
The temperature of the cool component was 1.3--1.6 keV, 
somewhat lower by 0.2--0.3 keV than
that in the fitting with the single-temperature model.
Since the core radius of the X-ray surface brightness is 
about 35$''$, the spectra within 40$''$ are thought 
to be affected by a projection effect.
However, the purpose of our work is to search for evidences 
of jet heating from relative temperature variations. 
The projection effect on the best-fit temperature is 
at most 0.2--0.3 keV, which is not very significant 
for our purpose.
Therefore, here after, we discuss our results 
on the single-temperature model fitting.
Figure 4 shows radial profiles of $kT$ and $N_{\rm H}$
obtained from spectral fittings with the single-temperature 
MEKAL model.

We first refer to the temperature profile (figure 4a).
At 150$''$ from the cluster center, the temperature reaches 
$\sim3.3$ keV, which is somewhat higher than the outer-region 
temperature obtained with ASCA: 3.0$\pm0.1$ keV 
at $3'-8'$ radius (\cite{fukazawa98}). 
This might be an indication of a radial temperature decrease 
toward the outer region, as claimed by \citet{markevitch98}.
Moreover, the temperature declines toward the cluster center, 
as reported based on a ROSAT analysis (\cite{irwin95}). 
Our Chandra results present finer and clearer profiles.
In the simple cooling flow model, the temperature is predicted 
to decline steeply down to $<$ 1 keV at the center.
However, the actual temperature profile does not indicate 
such a steep decline; the gas temperature within $30''$ is 
almost constant at $\sim$1.5 keV.
The temperature of the cool component in the two-temperature model 
is also higher than 1 keV.
In the case of A 2199, the temperature at the center is reported 
to be $\sim$2.0 keV, which is fairly high for 
the simple cooling flow picture.
Interestingly, both clusters exhibit apparent differences of 
the temperature profiles. 
The low-temperature region in 2A 0335+096 is extended up to 
$\sim150$ kpc ($\sim150''$),
while that of A 2199 is up to only $\sim100$ kpc.
The temperature profile in the very central region 
within $30''$ is flatter for 2A 0335+096 than for A 2199.

Secondly, regarding the absorption profiles (figure 4b), 
in the central region, the absorption is 
$\sim2.8\times10^{21}$ cm$^{-2}$, which is consistent with 
the previous results of Einstein SSS (\cite{white91}), 
ROSAT PSPC (\cite{irwin95}), and ASCA (\cite{kikuchi99}).
Therefore, problems concerning the response matrix of ACIS-S 
in the lower energy band (\cite{plucinsky02}) 
are not significant for our results.
On the other hand, our results show that the absorption is 
almost constant up to 3$'$ ($\sim200$ kpc). 
This indicates that there is no central excess absorption, 
and therefore no evidence of a large amount of 
cool gas predicted from the cooling flow picture. 
From a 21 cm line observation (\cite{stark92}), 
the hydrogen column density is estimated to be
$\sim1.71\times10^{21}$ cm$^{-2}$, which is half of the value 
compared with that inferred from the X-ray absorption.
This discrepancy implies the possible existence of 
galactic interstellar gas with high metallicity toward 
the direction of 2A 0335+096.

Finally, we refer to the metal abundance. 
The metal abundance is almost constant around 0.3 solar 
with a small variation of $\sim$0.1 solar 
over the ACIS-S field of view.
This trend is consistent with the ASCA results of 
0.34$\pm$0.04 solar and 0.30$\pm$0.03 solar (\cite{fukazawa00}),
which were derived from spectral fittings for the inner 2$'$ 
and outer $3'-8'$, respectively.
A more detailed investigation on the metal abundance is 
beyond the scope of this paper.
In summary, we have obtained results on the spectral 
properties of 2A 0335+096 that are consistent with 
the previous ones in the outskirts of the cluster, 
and definitely more accurate than previously published 
in the central regions, where we have spatially resolved
the temperature and absorption distribution.

\subsection{Temperature Map}
In subsection 3.2, we mentioned that the central temperature 
is higher than that predicted from the cooling flow picture. 
This temperature profile can be explained if there is 
a certain heating mechanism against cooling, for example, 
radio jets or merging. 
Here, we consider any evidence of jet heating. 
To examine the relation between the azimuthal temperature 
variation and radio jets, we obtained temperature maps 
with $5''$ square grids in a $60''\times60''$ region around 
the cluster center.
The background spectra and responses were prepared 
as described in the previous section.
The single-temperature MEKAL plasma model was used again, 
and could fit the spectra well.
We also performed this analysis on A 2199, 
since the square region of the temperature map 
obtained by \citet{johnstone02} is somewhat 
large for the scale of radio jets, 
and we would like to compare the maps with the same 
and finer grid scales.

Figure 5 shows two-dimensional temperature maps. 
The typical error of the temperature in each grid are 10\% 
and 15\% for 2A 0335+096 and A 2199, respectively.
Both clusters exhibit an asymmetric temperature structure.
In 2A 0335+096, although the radial temperature profile 
exhibits a gradual decrease toward the cluster center, 
as shown in figure 4a,the temperature structure 
around the central $60''\times60''$ region is fairly complex; 
there is no clear trend of circular symmetry.
There are two hot regions of $5''-10''$ size along 
the jet direction (NE and SW), located at $15''$ from the center, 
where the temperature is somewhat higher by 0.3--0.5 keV 
than that in the surrounding region.
A significant cool region is largely extended toward the SE region, 
and another local cool region exists at the NW direction.
Both cool regions are located perpendicular to the jets, and 
their temperature is $\sim1.3$ keV, lower than 
the central temperature of $\sim1.6$ keV.
We have found no evidence of a spectral hard component 
from jets or AGN.
For A 2199, the overall temperature structure is consistent 
with that of \citet{johnstone02}, and the circular symmetry 
is relatively better than that of 2A 0335+096.
The most noticeable cool region exists at the southern region 
of the center, and there is a weak hint of a higher temperature 
along the E and W jet directions.

In order to visualize the temperature variation more clearly 
and show the significance with error bars, 
we obtained the azimuthal temperature profiles with 
an azimuthal angle interval of 22$^\circ$.5.
The azimuthal temperature profiles were investigated 
on four concentric circles.
At each azimuthal angle, the spectrum was integrated within 
a circle whose radius was chosen to be 
5$''$, 5$''$, 10$''$, or 15$''$ 
on each concentric circle of 
10$''$, 20$''$, 40$''$ or 70$''$ from the 
cluster center, respectively.
The obtained azimuthal temperature profiles are 
shown in figure 6.
Here, the azimuthal angle is defined as 0$^\circ$ 
for the northern direction from the cluster center, 
and increases clockwise. 
In 2A 0335+096, the central radio jets lie at 135$^\circ$ 
(NE: northeast) and 315$^\circ$ (SW: southwest). 
It is clearly seen that the azimuthal temperature profile 
exhibits a large variation, which has been already noted 
by figure 5a.
The high-temperature region is seen at 320$^\circ$ 
for the inner region, which is almost the same direction 
as the SW jet. 
On the other hand, temperature depression exists around 
$200^\circ-260^\circ$ at the outer region, 
which is perpendicular to jets.
In A 2199, the radio jets lie at 90$^\circ$ (E: east) 
and 270$^\circ$ (W:west). 
A faint temperature depression is scarcely seen 
in the outer region 
around the E and W jets, as hinted in figure 5b.

As a result, some correlations between the temperature variation 
and radio jets are detected for both clusters, and 2A0335+096 
exhibits a larger amplitude of the temperature variation.
Therefore, we compared the radial temperature distributions 
between the jet direction and its perpendicular one, so that 
we can trace the region size of the temperature disturbance.
The resulting temperature distributions are shown in figure 7.
It can be seen that the profile is asymmetric.
The NE jet (open circles in the left region of figure 7)
shows a remarkably higher temperature at $10''-30''$ from the center 
than the opposite (SW), consistent with figure 6a.
The SE temperature-depression region in figure 6a 
(around $200^\circ-240^\circ$) is also clearly noticed at $20''-50''$ 
from the center in the left region of figure 7 (filled squares).

In summary, regions of both high and low temperature are found 
at $10''-40''$ from the center along the NE jet and the SE region 
for 2A 0335+096, 
while A 2199 exhibits an asymmetric temperature distribution 
in the direction perpendicular to the jets within 30$''$ 
of the center, and has a low-temperature region 
at the end of the E and W jet.

\section{Discussion}
We analyzed Chandra data for 2A 0335+096 and A 2199 to obtain 
the spatially resolved central properties of the ICM.
The X-ray surface brightness is not circularly symmetric 
at the center of both clusters. 
In 2A 0335+096, the SE part is brighter and a radio-quiet X-ray hole
exists at the NW part, although we find no significant disturbance by
radio jets.
On the other hand, a clear correlation with the East--West radio jets
is seen in A 2199; an X-ray emitting hot plasma is displaced by jets.
The temperatures of both clusters gradually decrease toward the center.
However, the central temperatures of both clusters are not 
as low as $<1$ keV; 1.6 keV and 2.0 keV for 2A 0335+096 and A 2199, 
respectively.
An asymmetric temperature distribution is found around the radio jets 
and the central $\sim30$ kpc region perpendicular to the jets.
It is interesting that the temperature variation is milder for A 2199, 
which exhibits a clearer disturbance of the ICM density by jets.
The radial profile of the absorption column density is fairly uniform, 
except for the central region within 50 kpc where some increases of 
$\sim10^{20}$ cm$^{-2}$ are barely seen.

The cooling time at the radius where temperature starts to drop is
2$\times10^{10}$ yr at 160$''$ (163 kpc) for 2A 0335+096 and 
$\sim9\times10^{9}$ yr at 100$''$ (89 kpc) for A 2199,
based on our Chandra results and \citet{johnstone02}.
Therefore, it can be said that the low-temperature region exists 
within the cooling radius, which is defined as the radius where 
the cooling time is comparable to the age of the cluster 
($\sim$10$^{10}$ yr). 
This result seems to be consistent with the cooling flow picture.
However, in this scenario, 
the action of radiative cooling leaves the central ICM to 
temperatures below 1 keV, values that we did not observe 
in the clusters under examination.
In particular, a high temperature of $\sim$2 keV 
at the center of A 2199 is not easy to explain. 
Therefore, the Chandra results for both clusters are not 
consistent with the simple cooling flow picture.
Our results indicate that although radiative cooling may occur, 
some heating mechanisms would prevent the ICM 
from cooling down to $kT<1$ keV.

An excess of the absorption column density toward 
the center of cooling-flow clusters is expected 
if a large amount of gas cools down 
and deposits at the cluster center. 
In the past, observations with the Einstein SSS reported 
the existence of absorbing cold matter (\cite{white91}), 
whose column density is (5--10)$\times10^{20}$ cm$^{-2}$; 
also, ROSAT observations sometimes claimed an increase 
in the absorption column density toward the cluster center
(e.g. \cite{allen97}). 
Einstein SSS observations reported that
2A 0335+096 exhibits a significant excess of 
the absorption column density: $3.0\times10^{21}$ cm$^{-2}$ 
against the galactic value of $1.7\times10^{21}$ cm$^{-2}$.
ROSAT PSPC observations also confirmed this value at the center 
of 2A 0335+096, and gave a hint of a gradual decrease of 
the absorption toward the outer region (\cite{irwin95}).
However, the narrower energy band and poorer energy resolution 
of Einstein and ROSAT compared with the Chandra observatory 
did not allow us to constrain the absorption column density 
accurately, because of the complex spectral properties 
at the cluster center.
Chandra ACIS-S, for the first time, determined the absorption 
column density accurately, as well as the radial distribution 
of the absorption, thanks to its energy resolution, 
wide energy band down to 0.3 keV, and superior angular resolution.
As a result, there is no sign of significant excess in absorption 
for both clusters.
A little evidence of excess absorption of $\sim1\times10^{20}$ 
cm$^{-2}$ within 50 kpc of the center in both clusters can 
be explained by the cold interstellar medium of mass 
$\sim10^9 M_{\odot}$ in the central galaxy.
For 2A 0335+096, the absorption value, itself, 
at the cluster center is consistent with the previously 
observed results, and the radial decrease that was 
previously suggested was not observed by ACIS-S within 
$\sim200$ kpc of the center.
Even if the distribution of cold material extends 
over $\sim200$ kpc, the absorption must be the largest 
toward the cluster center.
Therefore, our results rule out the existence of 
a large amount of cold material.
It can be said that the ROSAT PSPC results are consistent 
with no absorption gradient, considering their large errors 
(\cite{irwin95}).
However, a significant discrepancy remains that the obtained 
absorption is larger than the galactic value determined 
from radio observations.
We suggest that unidentified molecular clouds in our Galaxy 
exist toward the direction of 2A 0335+096.

We obtained a fine temperature map around the cluster center, 
and found that the temperature varies locally and mildly.
Both hot and cool regions are found along the radio jets, 
indicating that radio jets can certainly heat up the ICM locally 
and at the same time compress it to be adiabatically cooled.
Considering these results, it is suggested that radio jets 
recently ejected from cD galaxies were not powerful enough 
to give a large amount of shock-heated energy to the ambient ICM.
The temperature varies by about 1 keV on a scale of few tens 
of kpc at most with a gas density of $10^{-2}$ cm$^{-3}$, 
corresponding to work of $\sim10^{57}$ erg.
The cooling loss by radiation is estimated to be 
$10^{44}$ erg s$^{-1}$ at the center of both clusters.
We consider the effect of heat conduction, combined 
with the sound speed in the plasma, which reduces 
the temperature and density variation in the ICM.
The heat conduction flux was calculated to be 
$\sim10^{42}$ erg s$^{-1}$ for an $\sim1$ keV gradient 
in $\sim$10 kpc (\cite{spitzer56}). 
This heat flux shows that the typical heated-up regions 
found in both clusters vanish within $\sim10^7$ yr.
The crossing time scale of the sound velocity is also of 
the same order.
Hence, the ICM disturbance caused by jets can survive for 
as long as $\sim10^7$ yr.
The cumulative heating energy to reduce the radiative cooling 
over the past $10^7$ yr should have been $10^{59}$ erg, 
which is about 100-times larger than that inferred from 
the present ICM disturbance; in other words, 
the relatively mild variation of the ICM temperature and 
the density in both clusters implies that jet heating 
in both clusters is not effective enough to restrain 
the cooling flow, at least during the last $10^{7-8}$ years.
Therefore, cD galaxies should have ejected 
relativistic plasma and produced jets that are more powerful 
than that actually observed $>10^7$ years ago 
by an order of magnitude, if radio jets are attributed to 
a reduction of the radiative cooling, 
setting aside other heating processes, such as merging. 
A few nearby clusters have such a powerful radio jet 
at a redshift of $z<0.1$, like, e.g., the Cygnus-A cluster.
This implies that most cD galaxies might have ejected 
quite a powerful radio jet in the past, and such an active phase 
might have ended $10^9$ years ago. 

\bigskip
The authors thank an anonymous referee for a careful reading of the text 
and many helpful comments.
The authors are also grateful to the Chandra team for
their help in the spacecraft operation, calibration, and data analysis.

\onecolumn
\vspace{1.0cm}
\begin{table}[htbp]
\caption{Best-fit parameters from X-ray surface brightness 
fittings with the single $\beta$-model.}
\begin{center}
\begin{tabular}{ccccc}
 \hline
 \hline
   & $\beta$ & $r_{\rm c}$ & $Normalization$ & $\chi^2$\\
   &         & (arcsec)& ($10^{-3}$ count s$^{-1}$ arcsec$^{-2}$) & \\
 \hline
  2A 0335+096 & 0.58 & 32.1 & 1.26 & 1.87\\
\hline
\end{tabular}
\end{center}
\end{table}

\vspace{1.0cm}
\begin{table}[htbp]
\caption{$\chi^2$/dof for the one (1T) and two (2T) temperature models.}
\begin{center}
\begin{tabular}{ccccc}
 \hline
 \hline
 Radius & $\chi^2/dof$ for 1T & $\chi^2/dof$ for 2T\\
 \hline
 $0''-10''$  & 201.9/129 (1.56) & 155.3/128 (1.21)\\ 
 $10''-20''$ & 498.1/187 (2.66) & 385.0/186 (2.07)\\ 
 $20''-30''$ & 492.4/210 (2.34) & 346.7/209 (1.66)\\ 
 $30''-40''$ & 412.2/221 (1.86) & 357.7/220 (1.63)\\ 
 $40''-50''$ & 347.2/229 (1.52) & 310.0/228 (1.36)\\ 
 $50''-60''$ & 325.7/228 (1.43) & 297.4/227 (1.31)\\ 
\hline
\end{tabular}
\end{center}
\end{table}

\vspace{1.0cm}
\begin{figure}[htbp]
\begin{center}
\begin{minipage}[tbhn]{8.0cm}
\FigureFile(7.5cm,6.5cm){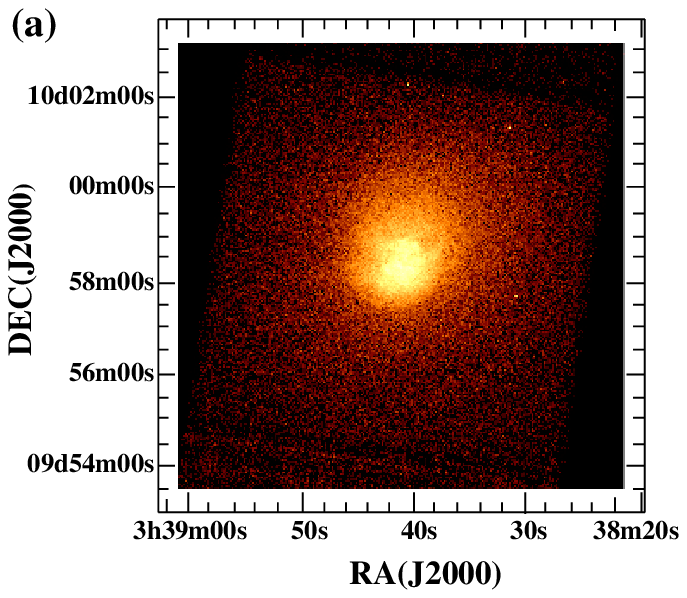}
\end{minipage}\quad
\begin{minipage}[tbhn]{8.0cm}
\FigureFile(6.7cm,6.0cm){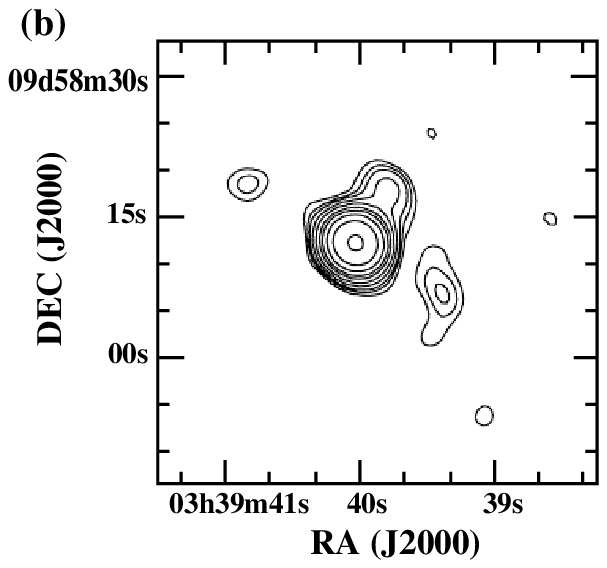}
\end{minipage}
\end{center}
\caption{Chandra ACIS-S3 X-ray image in the 0.4--9.0 keV band of
 2A 0335+096 (a), and 8.4 GHz radio image in the central 
$50''\times50''$region (b), which is quoted from \citet{sarazin95}.}
\end{figure}

\begin{figure}[htbp]
\begin{center}
\FigureFile(7.0cm,6.2cm){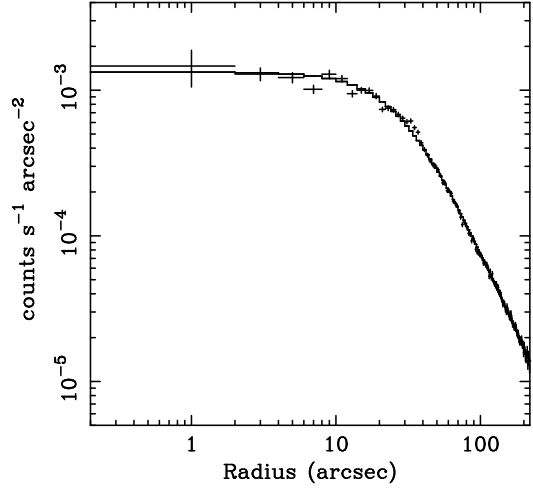}
\caption{X-ray surface brightness profile fitted with the $\beta$-model 
for 2A 0335+096.}
\end{center}
\end{figure}

\begin{figure}[htbp]
\begin{center}
\begin{minipage}[htbp]{8.0cm}
\FigureFile(7.5cm,6.2cm){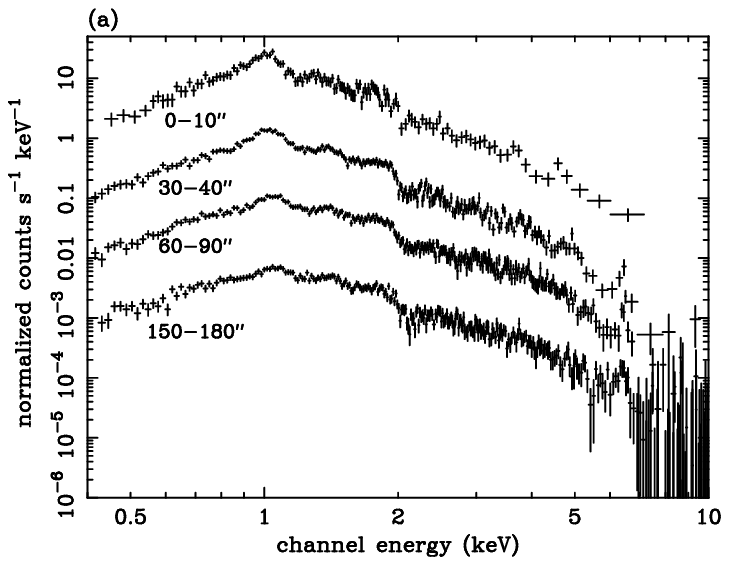}
\end{minipage}\quad
\begin{minipage}[htbp]{8.0cm}
\FigureFile(7.5cm,6.2cm){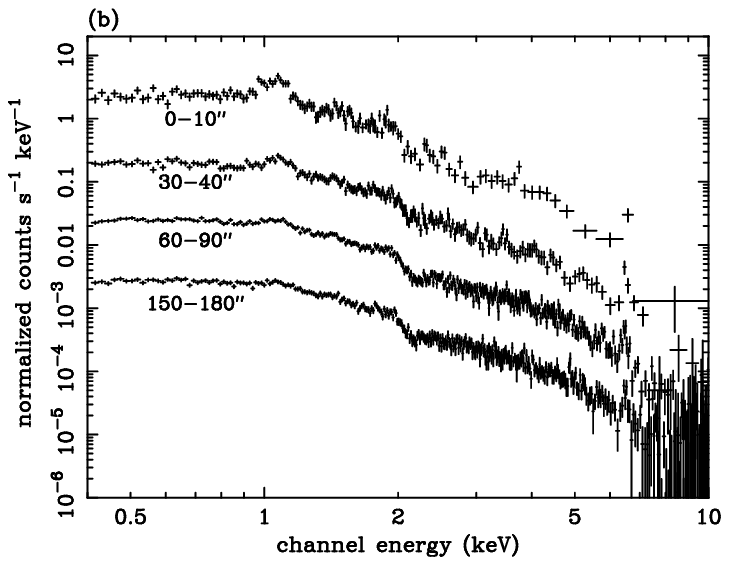}
\end{minipage}
\end{center}
\caption{Chandra ACIS X-ray spectra at four radii
of $0''-20''$, $60''-80''$, $120''-180''$ and $300''-360''$ 
centered on the cluster center.
(a) 2A 0335+096 and (b) A 2199. }
\end{figure}

\begin{figure}[htbp]
\begin{minipage}[htbp]{8.0cm}
\begin{center}
\FigureFile(7.5cm,6.2cm){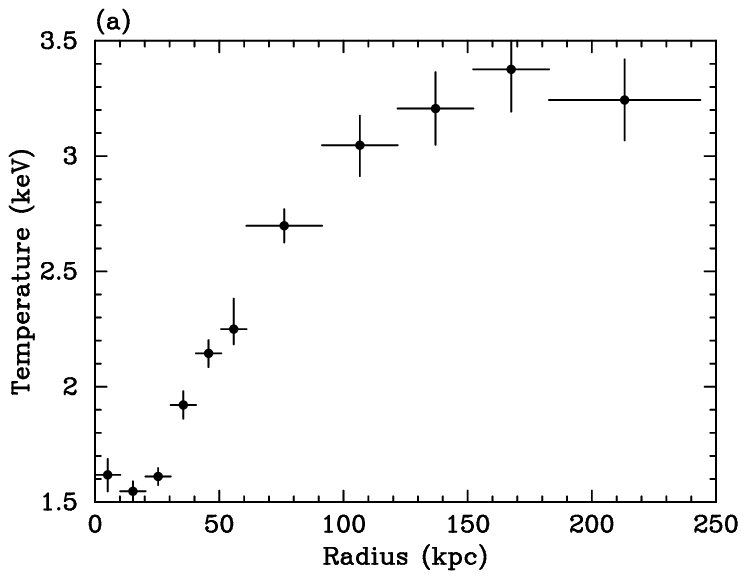}
\end{center}
\end{minipage}\quad
\begin{minipage}[htbp]{8.0cm}
\begin{center}
\FigureFile(7.5cm,6.2cm){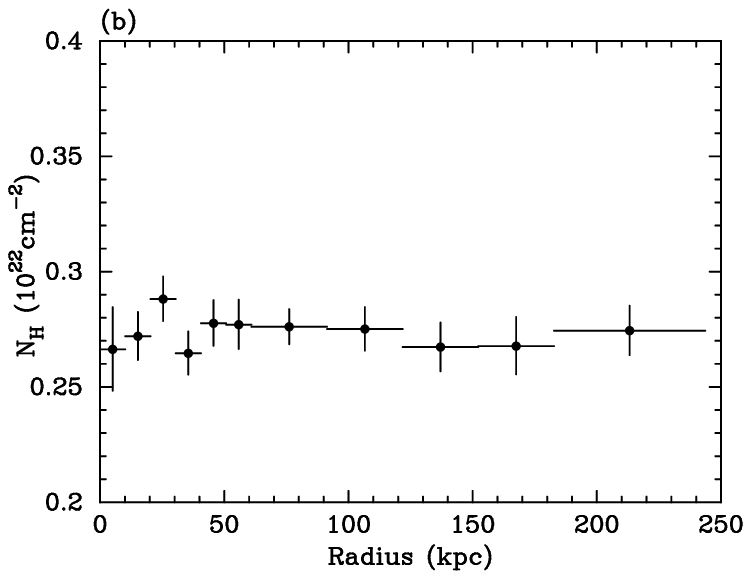}
\end{center}
\end{minipage}
\caption{Radial profiles of (a) the temperature, $kT$, and 
(b) the absorption column density, $N_{\rm H}$, for 2A 0335+096.}
\end{figure}

\begin{figure}[hbpt]
\begin{center}
\begin{minipage}[htbp]{8.0cm}
\FigureFile(7.5cm,8.0cm){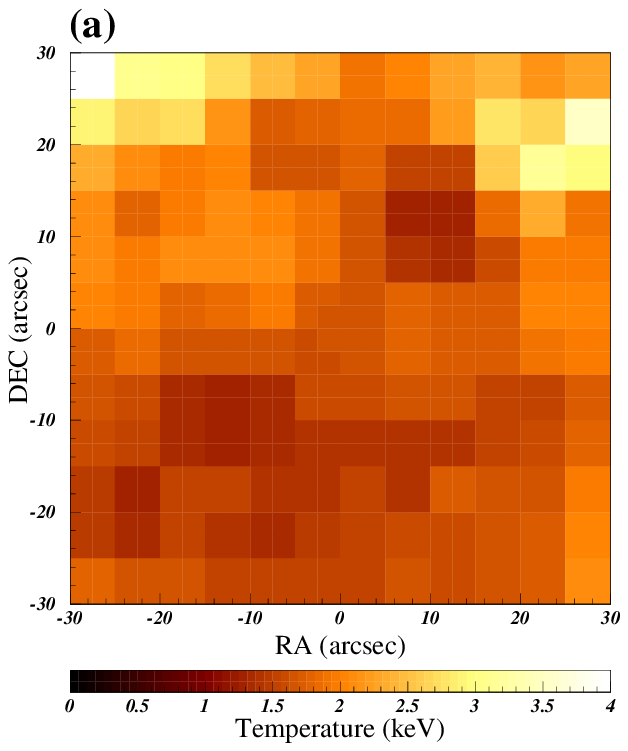}
\end{minipage}\quad
\begin{minipage}[htbp]{8.0cm}
\FigureFile(7.5cm,8.0cm){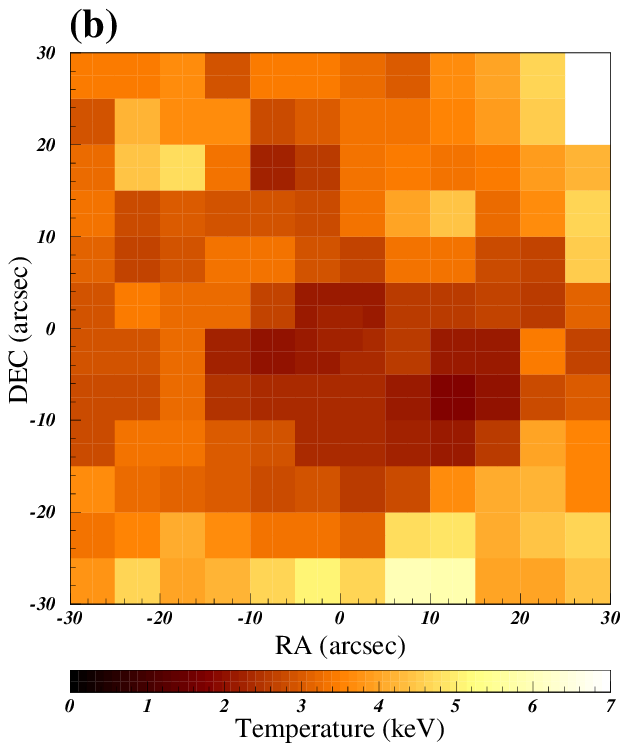}
\end{minipage}
\end{center}
\caption{Two-dimensional temperature maps of 2A 0335+096 (a) 
and A 2199 (b) in $60''\times60''$ around the cluster center. 
The darker regions correspond to lower temperature ones.
Typical errors of temperature in each grid are 10\% and 15\% for
2A 0335+096 and A 2199, respectively.} 
\end{figure}

\begin{figure}[htbp]
\begin{center}
\begin{minipage}[htbp]{8.0cm}
\FigureFile(7.5cm,6.2cm){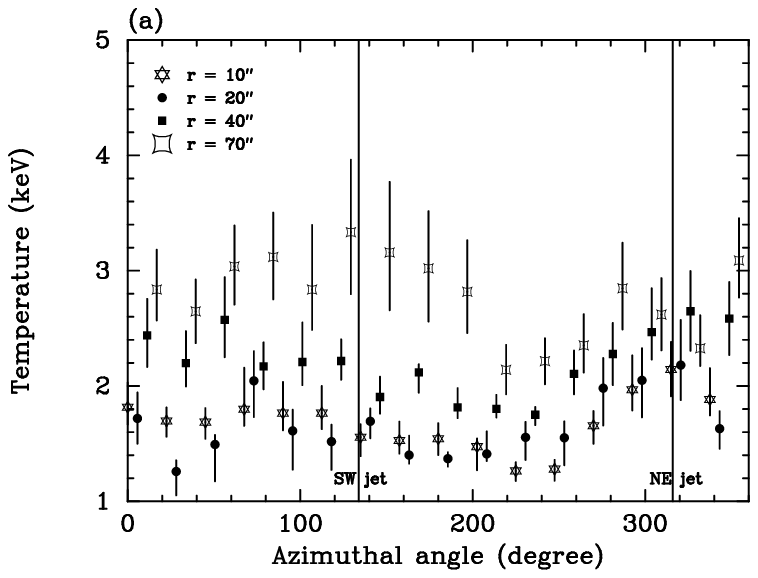}
\end{minipage}\quad
\begin{minipage}[htbp]{8.0cm}
\FigureFile(7.5cm,6.2cm){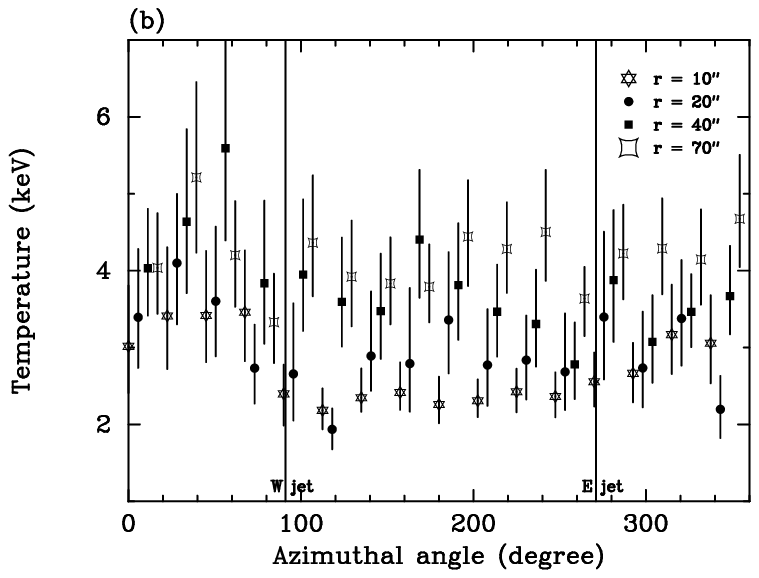}
\end{minipage}
\end{center}
\caption{Azimuthal temperature distributions of 2A 0335+096 (a) 
and A 2199 (b) on the four concentric annular regions, 
whose radii are 10$''$, 20$''$, 40$''$, and 70$''$ 
from the cluster center. The azimuthal angle is defined 
as 0$^\circ$ for the northern direction from the cluster center 
and increases clockwise.} 
\end{figure}

\begin{figure}[htbp]
\begin{center}
\FigureFile(7.5cm,6.2cm){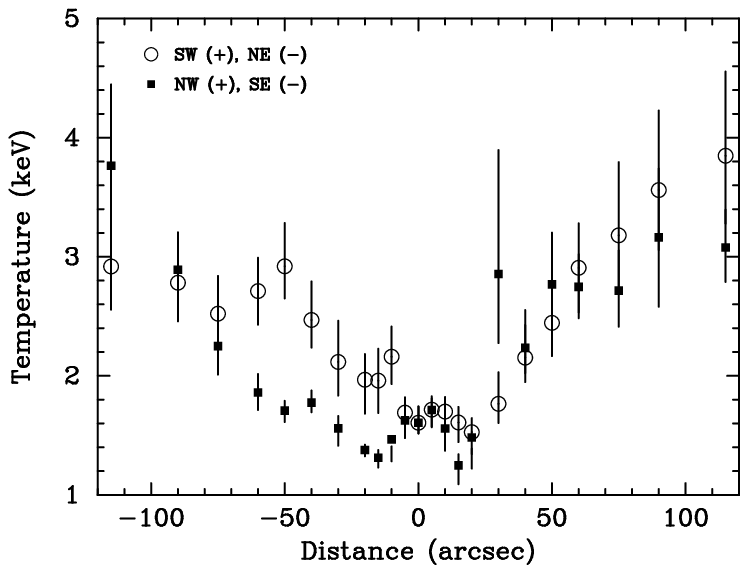}
\end{center}
\caption{Temperature distributions along the jet (open circles) and 
its perpendicular direction (filled squares) for 2A 0335+096. 
The horizontal axis is the distance from the cluster center 
in units of arcsec. Profiles along the SE (jet) and NE (perpendicular) 
directions are plotted in the positive region.}
\end{figure}


\begin{thebibliography}{}
\bibitem[Allen, Fabian (1997)]{allen97} 
Allen,~S.~W., \& Fabian,~A.~C. 1997, \mnras, 286, 583

\bibitem[Churazov et al. (2000)]{churazov00} 
Churazov,~E., Forman,~W., Jones,~C., \&  B\"ohringer,~H. 
2000, A\&A, 356, 788

\bibitem[David et al. (2001)]{david01} 
David,~L.~P., Nulsen,~P.~E.~J., McNamara,~B.~R., Forman,~W., Jones,~C., 
Ponman,~T., Robertson,~B., \& Wise,~M. 2001, \apj, 557, 546

\bibitem[Edge et al. (1992)]{edge92} 
Edge,~A.~C., Stewart,~G.~C., \& Fabian,~A.~C. 1992, \mnras, 258, 177

\bibitem[Ettori et al. (2002)]{ettori02} 
Ettori,~S., Fabian,~A.~C., Allen,~S.~W., \& Johnstone,~R.~M. 
2002, \mnras, 331, 635

\bibitem[Fabian (1994)]{fabian94} 
Fabian,~A.~C. 1994, ARA\&A, 32, 277

\bibitem[Fabian et al. (2000)]{fabian00} 
Fabian,~A.~C., et al. 2000, \mnras, 318L, 65

\bibitem[Fabian et al. (2001)]{fabian01} 
Fabian,~A.~C., Sanders,~J.~S., Ettori,~S., Taylor,~G.~B., 
Allen,~S.~W., Crawford,~C.~S., Iwasawa,~K., \& Johnstone,~R.~M., 
2001, \mnras, 321L, 33

\bibitem[Fukazawa (1997)]{fukazawa97} 
Fukazawa,~Y. 1997, Ph.D. Thesis, The University of Tokyo

\bibitem[Fukazawa et al. (1998)]{fukazawa98} 
Fukazawa,~Y., Makishima,~K., Tamura,~T., Ezawa,~H., Xu,~H., 
Ikebe,~Y., Kikuchi,~K., \& Ohashi,~T. 1998, \pasj, 50, 187

\bibitem[Fukazawa et al. (2000)]{fukazawa00} 
Fukazawa,~Y., Makishima,~K., Tamura,~T., Nakazawa,~K., 
Ezawa,~H., Ikebe,~Y., Kikuchi,~K., \& Ohashi,~T. 2000, \mnras 313, 21

\bibitem[Irwin, Sarazin (1995)]{irwin95} 
Irwin,~J.~A., \& Sarazin,~C.~L. 1995, \apj, 455, 497

\bibitem[Johnstone et al. (2002)]{johnstone02} 
Johnstone,~R.~M., Allen,~S.~W., Fabian,~A.~C., 
\& Sanders,~J.~S. 2002, \mnras, 336, 299

\bibitem[Jones, Forman (1984)]{jones84} 
Jones,~C. \& Forman,~W. 1984, \apj, 276, 38

\bibitem[Kikuchi et al. (1999)]{kikuchi99} 
Kikuchi,~K., Furusho,~T., Ezawa,~H., Yamasaki,~N.~Y., Ohashi,~T., 
Fukazawa,~Y., \& Ikebe,~Y. 1999, \pasj, 51, 301

\bibitem[Liedahl et al. (1995)]{liedahl95} 
Liedahl,~D.~A., Osterheld,~A.~L., 
\& Goldstein,~W.~H. 1995, \apj, 438L, 115

\bibitem[Markevitch et al. (1998)]{markevitch98} 
Markevitch,~M., Forman,~W.~R., Sarazin,~C.~L., 
\& Vikhlinin,~A. 1998, \apj, 503, 77

\bibitem[Makishima et al. (2001)]{makishima01} 
Makishima,~K., et. al. 2001, \pasj, 53, 401

\bibitem[Mazzotta et al. (2002)]{mazzotta02} 
Mazzotta,~P., Kaastra,~J.~S., Paerels,~F.~B., Ferrigno,~C., 
Colafrancesco,~S., Mewe,~R., \& Forman,~W.~R. 2002, \apj, 567, 37

\bibitem[McNamara et al. (2000)]{mcnamara00} 
McNamara,~B.~R., et al. 2000, \apj, 534L, 135

\bibitem[McNamara et al. (2001)]{mcnamara01} 
McNamara,~B.~R., et al. 2001, \apj, 562L, 149

\bibitem[Owen, Eilek (1998)]{owen98} 
Owen,~F.~N., \& Eilek,~J.~A. 1998, \apj, 493, 73

\bibitem[Peterson et al. (2001)]{peterson01} 
Peterson,~J.~R., et al. 2001, A\&A, 365, 104

\bibitem[Plucinsky et al. (2002)]{plucinsky02} 
Plucinsky,~P.~P., et al. 2002, astro-ph/0209161

\bibitem[Sarazin et al. (1995)]{sarazin95} 
Sarazin,~C.~L., Baum,~S.~A., \& O'Dea,~C.~P. 
1995, \apj, 451, 125

\bibitem[Sarazin et al. (1992)]{sarazin92} 
Sarazin,~C.~L., O'Connell,~R.~W., \&  McNamara,~B.~R. 
1992, \apj, 397L, 31

\bibitem[Schmidt et al. (2001)]{schmidt01} 
Schmidt,~R.~W., Allen,~S.~W., \& Fabian,~A.~C. 
2001, \mnras, 327, 1057

\bibitem[Schmidt et al. (2002)]{schmidt02} 
Schmidt,~R.~W., Fabian,~A.~C., \& Sanders,~J.~S. 
2002, \mnras, 337, 71

\bibitem[Spitzer (1956)]{spitzer56} 
Spitzer,~L.~Jr. 1956, Physics of Fully Ionized Gases
(New York: Interscience)

\bibitem[Stark et al. (1992)]{stark92} 
Stark,~A.~A., Gammie,~C.~F., Wilson,~R.~W., Bally,~J., Linke,~R.~A., 
Heiles,~C., \& Hurwitz,~M. 1992, ApJS, 79, 77

\bibitem[Tamura et al. (2001a)]{tamura01a} 
Tamura,~T., et al. 2001a, A\&A, 365L, 87

\bibitem[Tamura et al. (2001b)]{tamura01b} 
Tamura,~T., Bleeker,~J.~A.~M., Kaastra,~J.~S., 
Ferrigno,~C., \& Molendi,~S. 2001b, A\&A, 379, 107

\bibitem[White et al. (1991)]{white91} 
White,~D.~A., Fabian,~A.~C., Johnstone,~R.~M., 
Mushotzky,~R.~F., \& Arnaud,~K.~A. 1991, \mnras, 252, 72

\end{thebibliography}
\end{document}